\documentclass[aps,showpacs,twocolumn,prl]{revtex4-1}
\usepackage{graphicx}
\usepackage{amsmath}
\begin{document}
\title{Local Geometric Phase and Quantum State Tomography in a Superconducting Qubit}
\author{Kicheon Kang}
\affiliation{Department of Physics, Chonnam National University,
 Gwangju 500-757, Republic of Korea,}
\date{\today}
\begin{abstract}
We investigate quantum state reconstruction
of a superconducting qubit threaded by an Aharonov-Bohm flux, with 
particular attention to the local geometric phase.
A state reconstruction scheme is introduced with a proper account of the local
geometric phase generated by Faraday's law of induction.
Our scheme is based on measurement of three complementary quantities,
that is, the extra charge and two local currents.
Incorporating time-reversal symmetry and
 the Faraday's law, we show that the full density matrix
can be reconstructed without ambiguity in the choice of gauge.
This procedure clearly demonstrates that the
quantum Faraday effect plays an essential role in the dynamics of a
quantum system that involves Aharonov-Bohm flux.
%
\end{abstract}
\maketitle

Aharonov-Bohm (AB) effect~\cite{aharonov59} is regarded as a purely
topological phenomenon
that arises even in the absence of electromagnetic
force, as far as the magnetic field is localized inside a loop and
vanishes in the region of the electronic path~\cite{aharonov05}.
The situation is different if an AB loop involves a time-dependent
flux. Faraday's law of induction plays a central role in 
quantum state evolution in the presence of time-dependent magnetic
flux~\cite{kang12}, even in the adiabatic limit. The geometric
nature of the Faraday-induced phase has been investigated in Ref.~\cite{kang12},
which is essential in quantum state dynamics of generic AB loops.
It has been predicted that this phase is observable in a flux-switching
experiment with double-dot AB loop.
In addition, estimating the state of a qubit is
an essential ingredient for quantum information processing~\cite{nielsen00}.
The most developed candidate for a solid-state realization is the
superconducting
qubits~\cite{makhlin01}. 
AB flux is an essential control parameter in various types of superconducting 
qubits.  In particular, in a single Cooper pair
box (SCB) with two parallel-coupled Josephson junctions, the AB flux 
penetrating the loop
between the two junctions is used to control the effective
coupling strength of the two charge states~\cite{makhlin99}.
Also, flux switching may be useful for estimating the qubit
state in a SCB~\cite{liu05}. However, as pointed out previously by the 
author~\cite{kang12}, it is not possible to specify the state evolution
of a qubit involving a change in the AB flux, without proper
consideration of the Faraday's law of induction. It is rather puzzling that
this effect has been widely ignored.

In this Letter, we investigate how quantum state tomography 
(QST)~\cite{vogel89,paris04} can be
performed for a flux-tunable superconducting qubit, with particular attention
to the Faraday-induced local geometric phase. 
We introduce a QST scheme with detection of the local charge and the
two local currents flowing through each junction. The Faraday-induced
local phase plays an essential role in this procedure. Starting from a system 
with a time-reversal symmetry (that is, with a vanishing external magnetic 
field), the density matrix can be fully reconstructed without
ambiguity, and the phase evolution of its off-diagonal elements is 
completely determined by Faraday's law of induction.
Notably, this is in strong contrast with the arbitrary (gauge-dependent) 
local phase in the conventional description of the AB effect.
In addition, this procedure
provides particular insight for characterizing an equilibrium state.

{\em A superconducting Cooper pair box} -
We consider a superconducting Cooper pair box (SCB)
with two Josephson junctions threaded by a magnetic
flux (Fig.~1)~\cite{makhlin99}. This is one of the simplest quantum systems 
involving
the AB phase.
The SCB has been extensively studied in the context of quantum
information processing~\cite{makhlin01}, and is one of the best candidates
for investigating
the {\em Faraday-induced local phase}. A theoretical scheme
with a SCB~\cite{liu05} describes a state reconstruction with charge
detection followed by voltage/flux switching. However, the Faraday effect 
has not been considered
in Ref.~\onlinecite{liu05}. In fact, without considering the
Faraday effect,
the evolution of the qubit state cannot be properly described~\cite{kang12}.
As we will show here, characterization of the local phase, which is directly
related to Faraday's law of induction, is essential for a QST.
Despite recent progress in realizing the QST with superconducting
qubits~\cite{qst-sc-exp}, the essential role of the Faraday-induced local phase
has never been addressed.

A SCB (Fig.~1) is described by the Hamiltonian
\begin{subequations}
\label{eq:hamil}
\begin{equation}
 H = \sum_{n=0}^1 E_n |n\rangle\langle n|
   - \frac{1}{2} \left( \tilde{E}_J|1\rangle\langle 0|
                      + \tilde{E}_J^*|0\rangle\langle 1|   \right) .
\end{equation}
The qubit energy level $E_n$ ($n=0,1$) is $E_n = E_c(n-n_g)^2$, where
$E_c$ is the charging energy of a Cooper pair. This level is tunable
via the gate-dependent parameter $n_g$.
Josephson energy, $E_J$, is assumed to be
identical for the two junctions, which gives the effective Josephson coupling
\begin{equation}
 \tilde{E}_J = 2E_J e^{i(\varphi_a-\varphi_b)/2} \cos{(\varphi/2)} ,
\end{equation}
\end{subequations}
where $\varphi_a$($\varphi_b$) is the local phase shift across the junction
$a$($b$) (Fig.~1). The magnitude of $\tilde{E}_J$ is controlled by the
AB phase $\varphi
= \varphi_a+\varphi_b$,
whereas the choice of the two local phases $\varphi_a$ and $\varphi_b$ is
arbitrary. The phase factor $e^{i(\varphi_a-\varphi_b)/2}$ is
widely ignored for convenience, which is fine for describing any
phenomena with a time-independent flux. However, this phase factor
plays a major role in our context.
It is useful to rewrite the Hamiltonian in a
Bloch-sphere representation (assigning the pseudospin states
$|\uparrow\rangle = |0\rangle$, $|\downarrow\rangle = |1\rangle$) as
\begin{subequations}
\label{eq:hamil-bloch}
\begin{equation}
 H = -\frac{1}{2} \mathbf{B}\cdot\vec{\sigma} ,
\end{equation}
where
\begin{equation}
 \mathbf{B} = \left(\mathrm{Re}(\tilde{E}_J),\mathrm{Im}(\tilde{E}_J),
  E_c(1-2n_g)\right),
\end{equation}
\end{subequations}
and $\vec{\sigma}=(\hat{\sigma}_1,\hat{\sigma}_2,\hat{\sigma}_3)$.
As one can find from Eqs.(\ref{eq:hamil}, \ref{eq:hamil-bloch}),
the Hamiltonian depends on the phase factor
$e^{i(\varphi_a-\varphi_b)/2}$. Therefore, variation in the local phases affects
the quantum state dynamics, and one can already expect that the local phase
variation should be related to some physical process.

{\em Faraday-induced local phase} -
Any choice of $\varphi_a$ and $\varphi_b$ (with the constraint $\varphi_a+
\varphi_b= \varphi$) is fine unless a change
of the flux is involved.
However, when the flux varies in time, even in the adiabatic limit,
the Faraday's law of induction plays a crucial role
in the state evolution~\cite{kang12}. This should be considered
for the choice of gauge.
A change in the flux induces changes in the local phases $\varphi_a$ and
$\varphi_b$ as well as the AB phase $\varphi$, namely $\delta\varphi =
\delta\varphi_a + \delta\varphi_b$. For the SCB under consideration,
we adopt the representation with single-valued time-independent energy levels
$E_n$ ($n=0,1$)
for each qubit state. This is equivalent to the choice of time-independent
scalar potential. Then, we find
\begin{equation}
 \delta\varphi_{a(b)} = \frac{2e}{\hbar c}
  \int_{a(b)} \delta\mathbf{A} (\mathbf{r})\cdot d\mathbf{r}  ,
\label{eq:local-phase-qubit}
\end{equation}
where $\int_{a(b)}$ is the integral along the path $a(b)$.
The important point here is that the change in the vector potential
$\delta\mathbf{A}$ is not gauge dependent but is proportional
to the Faraday-induced momentum kick ($\mathbf{\delta p}$) as
\begin{equation}
 \delta\mathbf{A} = -c\int\mathbf{E}_t\, dt
   = -\frac{c}{e}\mathbf{\delta p},
\end{equation}
where $\mathbf{E}_t$ denotes the time-dependent contribution of the electric field generated by the Faraday's induction.
Therefore, $\delta\varphi_a$($\delta\varphi_b$) is a gauge-invariant 
physical quantity,
whereas $\varphi_a$($\varphi_b$) itself can be chosen arbitrarily to
describe the AB effect.

Basically, the {\em Faraday-induced local phase} is determined by the 
geometry of the system, and can be measured via flux switching and the charge
response of the qubit, as also described in Ref.~\onlinecite{kang12}
for a double-dot AB loop.
This is because the quantum dynamics of the qubit upon flux switching
is uniquely determined by Faraday's law of induction.

{\em Quantum state reconstruction procedure} -
Once (variation of) the local geometric phase is well defined due to the
law of Faraday induction, it is possible to carry out tomography of an 
arbitrary quantum state.
Three independent quantities should be measured 
for qubit state reconstruction; 
namely, the three components of the pseudospin average
($\langle\hat{\sigma}_1\rangle, \langle\hat{\sigma}_2\rangle,
\langle\hat{\sigma}_3\rangle$)
in the Bloch-sphere representation. This is easily understood
because any single-qubit
density matrix can be represented by~\cite{paris04}
\begin{equation}
 \rho = \frac{1}{2}\sum_{k=0}^3 \langle\hat{\sigma}_k\rangle \hat{\sigma}_k \,,
\label{eq:qst-spin}
\end{equation}
where $\hat{\sigma}_0 = \mathbf{1}$ denotes the unit matrix.
It is also possible to perform a QST
by measuring only one variable ($\langle\hat{\sigma}_3\rangle$, for
instance) followed by appropriate
single-bit operations. 
It can be done for a SCB with a charge detection
combined with pseudo-spin rotations, where the pseudo-spin rotation is 
performed by voltage and flux switching~\cite{liu05}. 
Here we introduce an
alternative, instructive rather than practical, approach.
Our scheme does not require the voltage or flux
switching necessary for single-bit operations. Instead, 
a series of direct measurements can be made for the three physical variables.
Note that the conclusion
we draw here does not depend on the kind of state-reconstruction scheme.
In any case, the Faraday-induced phase shift should be included,
which is indeed the essential factor determining the off-diagonal components 
of the density matrix.

Our scheme is based on direct measurement of the three complementary
quantities. For a SCB, the three complementary
variables correspond to the excess charge in the box and the two local currents
across each junction $a$ or $b$. The charge $\hat{q}$,
\begin{equation}
 \hat{q} = 2e\frac{\partial H}{\partial E_1} = 2e|1\rangle\langle1| ,
\end{equation}
gives the $z$-component of the pseudospin due to the relation
\begin{equation}
 \hat{\sigma}_3 = \mathbf{1} - \hat{q}/e.
\end{equation}
The local current $\hat{I}_\alpha$ flowing through junction $\alpha$ ($=a$ or $b$)
is given by
\begin{eqnarray}
 \hat{I}_\alpha &=& -\frac{2e}{\hbar}\frac{\partial H}{\partial\varphi_\alpha}
 \label{eq:ialpha} \\
   &=& -I_0 \left(\hat{\sigma}_1\sin{\varphi_\alpha}
                 \mp \hat{\sigma}_2 \cos{\varphi_\alpha}
            \right), \nonumber
\end{eqnarray}
where $I_0 = e E_J/\hbar$ is the current amplitude, and the sign $-$($+$) in the equation is for the case $\alpha = a$($b$).
This relation is derived from the fact that the local current density
 $\hat{\mathbf{j}}(\mathbf{r})$ of an arbitrary quantum system
 is obtained from the functional
derivative of the Hamiltonian with respect to the vector potential $\mathbf{A}$, as
\begin{equation}
 \hat{\mathbf{j}}(\mathbf{r}) = -c\frac{\delta H}{\delta\mathbf{A}(\mathbf{r})} \,.
\end{equation}
From Eq.~(\ref{eq:ialpha}), we find
\begin{subequations}
\label{eq:sigma12}
\begin{eqnarray}
 \hat{\sigma}_1 &=& -\frac{1}{I_0\sin{\varphi}} \left(
 \hat{I}_a \cos{\varphi_b} + \hat{I}_b\cos{\varphi_a} \right) ,   \\
 \hat{\sigma}_2 &=& \frac{1}{I_0\sin{\varphi}}  \left(
  \hat{I}_a\sin{\varphi_b} - \hat{I}_b\sin{\varphi_a}\right) .
\end{eqnarray}
\end{subequations}
The average values of the three quantities, $\langle\hat{q}\rangle$,
$\langle \hat{I}_a\rangle$, and $\langle \hat{I}_b\rangle$ provide
full information of the
average values of the three complementary variables, $\langle\hat{\sigma}_1\rangle$,
$\langle\hat{\sigma}_2\rangle$, and $\langle\hat{\sigma}_3\rangle$. Therefore, a
complete reconstruction for a given qubit state is possible from
Eq.~(\ref{eq:qst-spin}).

{\em Time-reversal symmetry (TRS) and the local phase} -
While $\langle\hat{\sigma}_3\rangle$ is uniquely determined by measuring excess
charge,  $\langle\hat{\sigma}_1\rangle$ and $\langle\hat{\sigma}_2\rangle$
depend on the choice of gauge $\varphi_a,\varphi_b$.
In the Bloch-sphere
representation, this is equivalent to the choice of the $x-y$ axes.
However, further constraints on the gauge can be provided by imposing the symmetry of the
system. For example, let us start a quantum state reconstruction procedure 
from the case with TRS. The TRS is achieved when the external
magnetic field is zero in all regions of the system.
The time-reversed Hamiltonian ($\bar{H}$) is related to the original Hamiltonian
of Eqs.~(\ref{eq:hamil},\ref{eq:hamil-bloch}) as
\begin{equation}
 \bar{H} = \bar{H}(\varphi_a,\varphi_b) = H(-\varphi_a,-\varphi_b)  .
\end{equation}
The TRS condition, $\bar{H}=H$, is satisfied by imposing the constraint
$\varphi_a=\varphi_b=0$~\cite{TRS}.
That is, for a system with TRS, the local phase is
uniquely determined, in contrast to an
arbitrary choice (with $\varphi=0$) for describing an AB loop~\cite{note-gauge}.
In fact, this constraint of the local phase is equivalent to the theorem:
the non-degenerate eigenfunction of a time-reversal invariant Hamiltonian
should be real (more generally, a real function times phase factor independent of
 position)~\cite{sakurai94}.

Starting from the case with perfect TRS, a complete state reconstruction 
procedure can be provided
as follows: (i) A quantum state $\rho$ (density matrix) is prepared in the absence
of the external magnetic field
($\varphi_a=\varphi_b=0$). (ii) The density matrix $\rho$ is reconstructed, without
ambiguity in the local phase, by measuring $\langle\hat{q}\rangle$,
$\langle \hat{I}_a\rangle$, and $\langle \hat{I}_b\rangle$, as described in
Eqs~(\ref{eq:qst-spin}-\ref{eq:sigma12}). This is possible
because the arbitrariness of the local phase is removed by TRS.
(iii) Magnetic field is turned on and varied, which
modifies the local phases by $\delta\varphi_a$ and $\delta\varphi_b$. Note that these
local phases are uniquely defined by the Faraday-induced momentum kick
 (Eq.~(\ref{eq:local-phase-qubit})).
(iv) The procedure of (i) and (ii) is repeated to provide a full 
reconstruction of
state $\rho$ as a function of the applied external magnetic field
($\varphi_a$ and $\varphi_b$ are uniquely given).
(v) In this way, an arbitrary state can be
reconstructed for an arbitrary distribution of the external magnetic field.

{\em Equilibrium-state reconstruction and persistent current} -
While the QST described above is valid for any state and is not limited to
an equilibrium, it is worth investigating the equilibrium state in relation to
the persistent current.
The standard definition of the persistent current of an AB loop is given 
by the derivative of the Hamiltonian with respect to the AB phase,
\begin{equation}
 \hat{I} = -\frac{2e}{\hbar}\frac{\partial H}{\partial\varphi} ,
\label{eq:pc}
\end{equation}
which leads to the relation
\begin{eqnarray}
 \hat{I} &=& -\frac{I_0}{2}
       \left(\sin{\varphi_a}+\sin{\varphi_b}\right)\hat{\sigma}_1
          + \frac{I_0}{2}
       \left(\cos{\varphi_a}-\cos{\varphi_b}\right)\hat{\sigma}_2
   \nonumber \\
   &=& \frac{1}{2}\left( \hat{I}_a + \hat{I}_b \right) ,
\end{eqnarray}
in our SCB.
This is an interesting expression in that the standard definition of the
persistent current is
equivalent to the average of the two local currents. In fact, the persistent
current of the definition
in Eq.~(\ref{eq:pc}) is meaningful only for an equilibrium state, whereas
the local currents $\hat{I}_a$ and $\hat{I}_b$ have their direct physical
meaning for any quantum state. This also implies that the conventional 
description
of the circulating persistent current is a limiting case of our 
local-current based scheme. In an equilibrium state, the two local
currents should be balanced,
$\langle\hat{I}_a\rangle
=\langle\hat{I}_b\rangle$. Naturally, it results in the obvious relation
$\langle\hat{I}\rangle =\langle\hat{I}_a\rangle=\langle\hat{I}_b\rangle$.

Further, by imposing this equilibrium condition $\langle\hat{I}_a\rangle
=\langle\hat{I}_b\rangle$, we find
\begin{equation}
 \langle\hat{\sigma}_2\rangle
    = \langle\hat{\sigma}_1\rangle \tan{\frac{\varphi_a-\varphi_b}{2}}.
\label{eq:s1s2}
\end{equation}
It can be shown by a straightforward evaluation that the qubit state is an incoherent
mixture of the two eigenstates $|+\rangle$ and $|-\rangle$
under the condition of Eq.~(\ref{eq:s1s2}).
That is, the density matrix is reduced to the form
\begin{equation}
 \rho = \rho_{eq} = a_+|+\rangle\langle+| + a_-|-\rangle\langle-| ,
\end{equation}
where $a_\pm$ satisfies $a_+ + a_- = 1$ with $0\leq a_\pm\leq1$.
This result is equivalent to the basic postulate of equilibrium
quantum statistical mechanics that the interference between different
eigenstates vanishes~\cite{huang87}. Interestingly,
this property is not postulated here but derived from equilibration of
the local current,
and can also be understood in terms of the stationary nature of
the eigenstates.

{\em Conclusion} -
In conclusion, the local geometric phase induced by Faraday's law of
induction plays a central role in quantum state tomography of a 
superconducting qubit
in the presence of an external magnetic flux.
A state reconstruction scheme has been proposed for a superconducting 
Cooper pair box
which involves a change in the flux. Together with the constraint of
time-reversal symmetry and the Faraday-induced local phase, any
quantum state can be reconstructed from measuring three complementary
quantities, without ambiguity in the gauge dependence. It is also important
to note that our conclusion is not limited to the specific case of a
superconducting qubit
but can be widely applied to any quantum system that involves a change in 
magnetic flux, which calls for further study.

This work was supported by the
National Research Foundation of Korea under Grant 
No.~2009-0084606, 2012R1A1A2003957, and by LG Yeonam Foundation.

%

\begin{figure}
\includegraphics[width=2.0in]{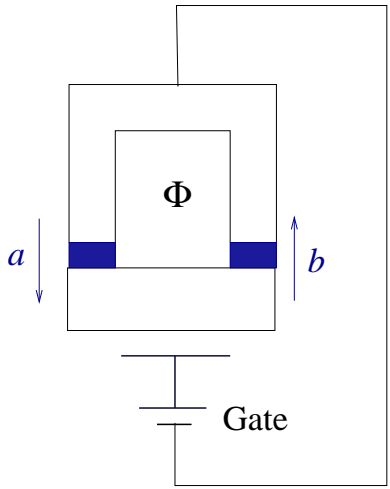}
\caption{(Color online) Schematic of a single Cooper pair box with two Josephson
 junctions ($a,b$) threaded by a magnetic flux $\Phi$. The qubit state is
 controlled by the gate voltage and
 the flux.
 }
\label{fig1}
\end{figure}


\begin{thebibliography}{99}
\bibitem{TRS} To be precise, $\varphi_a=-\varphi_b=\pi$ is an equally
possible choice satisfying both TRS and $\phi=0$. However, this is
equivalent to the case with negative $E_J$ and $\varphi_a=\varphi_b=0$.
\bibitem{note-gauge} Note that this is not in contradiction with the freedom of gauge choice (that gives an arbitrary choice of the local phase) in the standard description
of the Aharonov-Bohm effect with time-independent magnetic flux.
\end{thebibliography}
\end{document}